\documentclass[preprintnumbers,prd,aps,showpacs,amsmath,amssymb,floats,nofootinbib]{revtex4-1}
\usepackage{graphicx}
\usepackage{epsfig}
\usepackage{bm}% bold math
\usepackage{amsfonts,amssymb}
\usepackage{times}
\usepackage{natbib}
\usepackage[dvips]{color}

\def\lapp{\mathrel{\rlap{\raise.5ex\hbox{$<$}}
                    {\lower.5ex\hbox{$\sim$}}}}
\def\gapp{\mathrel{\rlap{\raise.5ex\hbox{$>$}}
                    {\lower.5ex\hbox{$\sim$}}}}
\def\com#1{{}^D\!C_{#1}}
\def\comp#1{{}^{D-1}\!C_{#1}}
\def\comq#1{{}^{D+1}\!C_{#1}}

\begin{document}
\title{Inflation in Higher Dimensional Gauss-Bonnet Cosmology}
\author{Isha Pahwa}
\email{ipahwa@physics.du.ac.in}
\affiliation{Department of Physics and Astrophysics, University of Delhi,
Delhi 110007, India.}

\author{Debajyoti Choudhury}
\email{debchou@physics.du.ac.in}
\affiliation{Department of Physics and Astrophysics, University of Delhi,
Delhi 110007, India.}

\author{T. R. Seshadri}
\email{trs@physics.du.ac.in}
\affiliation{Department of Physics and Astrophysics, University of Delhi,
Delhi 110007, India.}

\date{\today}

\begin{abstract}

A Gauss-Bonnet term naturally appears in the action for gravity when
one considers the existence of space time with dimensions more than
1+3. A variety of inflationary models can be obtained within such a
framework, once the scale factor for the hidden dimension(s) is
not constrained to be the same as that of the visible ones.  In
particular, the need for an adhoc inflaton field is eliminated.  
The phase space has a rich
structure with different types of solutions, both stable and
unstable. For a large class of solutions, the scale factors rapidly
approach an asymptotic exponential form.  Furthermore, sufficient 
inflation can be obtained for only a modest compression of the hidden 
world, if the latter is of a  sufficiently large dimension.
\end{abstract}
\maketitle
\section{Introduction}
The prospect of gravity propagating in a space of dimensions larger
than the canonical $(1+3)$ has long been an intriguing one. Apart from
being a natural generalization of General Relativity, the idea has
proffered tantalizing visions of both a way out of many vexing issues
in high energy physics as well as that of unification of gravity
with other forces. Starting with the efforts of Kaluza and Klein\cite{KK1,KK2},
passing through the formulation of theories in both large extra
dimensions as well as small but warped extra dimensions, to gravity
theories being duals to strongly coupled quantum field theories, it
has been a riveting saga. Added to this has been the promise of String
Theory as a path to both a consistent quantum theory of gravity as
well as unification with the strong and electroweak forces.

Naturally, this has also lead to the formulation of cosmology in
space-times with a larger number of dimensions, and the literature
abounds with a number of such scenarios. These encompass both models
wherein the visible world is confined to the usual $(1+3)$-dimensions,
viz. the braneworld models\cite{RS1,RS2,ADD,maeda}, as well as those where we traverse the
entire gamut of extra dimensions (with the latter, perforce, now being
compactified with a sub-attometer radius).

Whatever be the details of such models, it is quite apparent that the
presence of such extra dimensions would make themselves manifest at
suitably high energies, for the extra degrees of freedom associated
could then be excited. Verily, this is the guiding principle for both
collider searches as well as the study of their role in quantum
corrections to low-energy processes.  In a similar vein, their role in
the dynamics of the universe as a whole would be more pronounced when
the available energy was large, or equivalently when the distance
scales were small, in other words, the era of the early universe.

If we contemplate the quantum nature of gravity, it is almost
self-evident that the familiar Einstein equations cannot represent the whole
story.  Even if one started with the Einstein-Hilbert action at the
classical level, quantum corrections would generate higher derivative
terms in the effective action. The form of the latter, at first
glance, is protected only by the symmetries of the theory, which in
the present case, is general covariance. While this requirement, by
itself, is not very restrictive, one may treat the additional terms as
a power series in curvature (or, rather, in scalars formed out of the
curvature tensor) and, in a phenomenological treatment, confine
oneselves to the lowest non-trivial terms in this expansion. Even
then, three different terms are possible. One particular combination
viz. the Gauss-Bonnet term\cite{Lovelock,naresh,madore}, is of particular interest both on account
of certain interesting properties to be discussed below and for being
the lowest order correction to the Einstein-Hilbert action as
evaluated within String Theory\cite{boulware}. In terms of the Riemann tensor 
$R_{\mu\nu\phi\zeta}$, the Ricci tensor $R_{\mu\nu}$ and the Ricci scalar $R$, 
it can be expressed as 
\begin{equation}
 \textit{G} \, = \, R^2 \, - \,4 R_{\mu\nu} \, R^{\mu\nu} \, + \,
    R_{\mu\nu\phi\zeta} \, R^{\mu\nu\phi\zeta} \ .
\end{equation}
Interestingly, in $(1+3)$-dimensions, this is but a topological term and 
adds little to geometrodynamics. In a space of higher dimensions, this 
is no longer the case and the addition of this term to the 
action can lead to non-trivial 
modifications to the equations of motion, thereby giving rise to interesting 
effects. In this paper, we will analyze some of the features of such a theory
in context of the early universe.

In section 2, we look at the evolution equations and the conservation
equation, following by section 3 which lists all the cosmological
solutions and discusses the behaviour of the scale factors for both
the visible and the extra dimensions. In the end, i.e., section 4, we summarize
the paper and give conclusions.

\section{Evolution Equations}
The Einstein equations, as derived for the Einstein-Hilbert action
are second-order in the derivatives of the dynamical variable,
viz. the metric. This is true for any dimension of spacetime.
Since the dynamics of the Universe today is well explained by
this theory, it is paramount that any deviation from the 
Einstein-Hilbert action should essentially disappear 
in the context of the present-day universe atleast to first order.
In particular, this motivates the thought that even for higher dimensional
theories, the equations of motion need to be second order\footnote{While 
this requirement is not a strict one, it is certainly an useful one,
and also serves to eliminate ghosts, which are generic in higher
dimensional theories.}.
It restricts the form of the action and the simplest term that
one can write beside the Einstein-Hilbert and cosmological constant
term is the Gauss-Bonnet term. Such a deformation has the further advantage 
in the cosmological context that on account of it being a 
quadratic in the curvature, it could have played a significant
role in the early history of the universe, while its effect 
today would be negligible. 

We consider the higher dimensional Gauss-Bonnet action\cite{andrew,toporensky} as
\begin{equation}
 S \, = \, \int d^{4+D}x \sqrt{-g} \, 
   \left[ {\cal L}_m
    - \frac{M^{D+2}}{2}(\lambda R\, + \epsilon \, \textit{G}) \right]
 \label{action}
\end{equation}
where $R$ and $g$ correspond to the $(1+3+D)$ dimensional curvature scalar
and determinant of the metric tensor respectively.  Although we have not 
added a cosmological constant term explicitly, we can always if intended,
do so by adding an appropriate piece in the `matter' lagrangian $\cal L$.
$M$ defines the
scale of quantum gravity in the entire bulk, and is related to
$M_{\rm Planck}$ (defined in $4$-dimensions) through $V_D M^{D+2}\, = \,
M^2_{\rm Planck}$, where $V_D$ is volume of the extra-dimensional
subspace.  One would expect that $V_D \gtrsim M^{-D}$ and, thus $M
\lesssim M_{\rm Planck}$.  The quantity $\epsilon$ parametrizes the weight
of the Gauss-Bonnet term and can be reexpressed as $\epsilon \, = \,
\gamma M^{-2}$ where $\gamma$ is a dimensionless constant. If the
Gauss-Bonnet term is the result of quadratic corrections, then one
would typically expect $|\gamma|<1$. The two choices $\lambda =
+1(-1)$ correspond to attractive(repulsive) gravity in the Newtonian
limit. While $\lambda = -1$ is not unreasonable in an epoch where
the Gauss-Bonnet term dominates, clearly it is untenable at late
times (i.e., the present epoch) where the curvature is small and the
standard Einstein-Hilbert action needs to be recovered.  We shall,
then, assume $\lambda=+1$. It should be noted that 
String Theory predicts that $\gamma <0$, a result\cite{boulware,paul} that we shall adopt.

What we can demand purely from observations is the homogeneity and isotropy
of the normal $3-$dimensional space. If this is the only constraint to be 
satisfied, the line elment can be expressed as 
\begin{equation}
 ds^2 = -dt^2+a^2(t,x^I)\left(\frac{dr^2}{1-k_1 r^2} + r^2
 d\Omega\right) + 2g_{0I}dt dx^{I} + g_{JK}dx^Jdx^K .
\end{equation}
We will, however, resort to a simplification in which the hidden space
is also homogeneous and isotropic.  Neither the observed homogeneity
and isotropy nor any other conclusion about observable cosmology that
we shall draw is contingent on this assumption. On the other hand, this
serves to constrain the degrees of freedom in the theory, thereby
enhancing its falsifiability.  The scale factor $a(t)$ will, then, be
a function only of time.  Similarly, the extra dimensional sector of
the line element can also be expressed in a form similar to the normal
dimension with its own scale factor, $b(t)$.  While both these
sub-spaces are isotropic among themselves, the overall space-time is
manifestly anisotropic. This feature is built into the formalism by allowing
$a(t)$ and $b(t)$ to be different. Such a space-time which has one
temporal dimension, $3$ `visible' spatial dimensions and $D$ extra
spatial dimensions, can be described by the line element,
\begin{equation}
 ds^2 = -dt^2+a^2(t)\left(\frac{dr^2}{1-k_1 r^2} + r^2
 d\Omega\right)+b^2(t)\left(\frac{d R^2}{1-k_2 R^2}+R^2 d\Omega_{D-1}
\right) .
\label{lineelement}
\end{equation}
We denote the time dimension with super(sub)script `$0$'.
Whereas we reserve lower-case Roman indices ($i,j = 1,2,3$) for the normal 
spatial dimensions, upper case Roman indices denote
the extra dimensions and take values $I,J=4,5,\dots, D+3$.
The observable universe is well described 
by a vanishing spatial flatness and hence $k_1 = 0$. For the sake of 
simplicity, we likewise 
assume the extra dimensional curvature to be vanishing\footnote{Note that
$k_2\neq 0$, for example, $M^4\times S^D$ would, typlically, require
non-trivial energy-momentum tensor.},  i.e. $k_2=0$. In other words, 
our spacetime manifold is described by $M^4\times T^D$.

Starting from equation (\ref{action}), the Euler-Lagrange equation
for the metric would read
\begin{equation}
  \tilde{G}^{\mu}_{\nu} = \frac{1}{M^{2+D}} \, T^{\mu}_{\nu}
\end{equation}
where $T^{\mu}_{\nu}$ is the $1+3+D$ dimensional stress-energy tensor for the matter
content, and the generalised Einstein tensor $\tilde{G}^{\mu}_{\nu}$
incorporates the contribution from the Gauss-Bonnet term. Its
non-zero components are given by
\begin{eqnarray}
 \tilde{G}_0^0 &=& 
- \lambda \left[ 3 \, \frac{\dot{a}^2}{a^2} 
                + 3 \, D\frac{\dot{a}}{a}\frac{\dot{b}}{b}
                + \com{2} \frac{\dot{b}^2}{b^2} \right]
      + \epsilon \, \left[D \, \frac{\dot{a}^3 \dot{b}}{a^3 b} 
                         +36 \, \com{2} \, \frac{\dot{a}^2 \dot{b}^2}{a^2 b^2} 
                         +18 \, \com{3} \, \frac{\dot{a} \dot{b}^3}{a b^3} 
                          + 12 \, \com{4} \frac{\dot{b}^4}{b^4}  \right]
\label{Eeq00}
\\
\tilde{G}_i^i  &=& 
                   4 \, \epsilon \, \left[ 2 \,D\frac{\ddot{a}\dot{a}\dot{b}}{a^2b}
                   + 2 \, \com{2} \, \frac{\dot{b}^2\ddot{a}}{b^2a}
                   + D \, \frac{\dot{a}^2\ddot{b}}{a^2b}
                   + 3 \, \com{3} \, \frac{\dot{b}^2\ddot{b}}{b^3} 
+ 3 \, \com{2} \, \frac{\dot{a}^2\dot{b}^2}{a^2b^2}
                   + 6 \, \com{3} \frac{\dot{b}^3\dot{a}}{b^3a}
                   + 3 \, \com{4} \, \frac{\dot{b}^4}{b^4}
                   + 4 \, \com{2} \frac{\ddot{b}\dot{b}\dot{a}}{b^2a}\right]
\notag \\
               & - & \lambda \left[ 2 \,\frac{\ddot{a}}{a} 
                   + D \, \frac{\ddot{b}}{b} 
                   + 2 \,D \frac{\dot{a}}{a}\frac{\dot{b}}{b} 
                   + \frac{\dot{a}^2}{a^2} 
                   + \com{2} \frac{\dot{b}^2}{b^2} \right]
\qquad 
\forall \ i 
\label{Eeqii}\\
%%%
\tilde{G}_I^I  &=&  -\lambda \, \left[ (D-1) \frac{\ddot{b}}{b}
                    +3 \, \frac{\ddot{a}}{a}
                    +3 \, (D-1) \, \frac{\dot{a}}{a}\frac{\dot{b}}{b}
                    + \comp{2} \, \frac{\dot{b}^2}{b^2} 
                    +3 \, \frac{\dot{a}^2}{a^2} \right]
\nonumber \\
& + & 
    12 \, \epsilon \left[2\,(D-1) \, \frac{\ddot{a}\dot{a}\dot{b}}{a^2b}
                    +\comp{2} \, \frac{\dot{b}^2\ddot{a}}{b^2a} 
+(D-1) \frac{\dot{a}^2\ddot{b}}{a^2b}
                    +\comp{3} \frac{\dot{b}^2\ddot{b}}{b^3}
                    +3 \, \comp{2} \, \frac{\dot{a}^2\dot{b}^2}{a^2b^2}
\right. \notag \\
               &  & \left.   \hspace*{2em}  
                    +3 \, \comp{3} \, \frac{\dot{b}^3\dot{a}}{b^3a} 
                    + \comp{4} \, \frac{\dot{b}^4}{b^4}
                    +2 \, \comp{2} \, \frac{\ddot{b}\dot{b}\dot{a}}{b^2a} 
+ \frac{\dot{a}^2\ddot{a}}{a^3}
                    + (D-1) \, \frac{\dot{a}^3\dot{b}}{a^3b} \right]\hspace{1cm} \forall \ I \label{EeqII}
\end{eqnarray}
While the Gauss-Bonnet term gives rise to 
terms quadrilinear in the scale factors ($a(t)$ and $b(t)$) 
or their derivatives, they still contain only upto the second derivatives. 
(This feature is not particular to the cosmological solution, but is also 
retained in the general case.)
It is worth noting that, for $D=1$,
 eq.(\ref{EeqII}) is independent of $b(t)$ while this is not true for 
 $D>1$. This distinction is of profound importance 
and leads to some very interesting dynamical features that some earlier 
efforts\cite{sami}, concentrated as they were on $D=1$, had missed.
Indeed, the addition of the sixth, seventh and eigth dimensions, each 
brings into play a new term in the dynamics. 
Adding further to $D$ does not introduce any new term; 
instead, it would alter both 
the overall weight of the Gauss-Bonnet term relative to the 
Einstein-Hilbert term, and also subtly shift the balance between the 
various ${\cal O}(\epsilon)$ terms in the equations above, thereby 
affecting the dynamics. 

Note that our assumption of a metric of the form of equation
(\ref{lineelement}), or in other words, homogeneity and isotropy of
the universe separately in the visible and the extra dimensional
subspaces require that the stress-energy tensor be of the form 
\begin{equation}
 T^{\mu}_{\nu} = diag{(-\rho,P_a,P_a,P_a,P_b,....,P_b)}   \label{Tmunu}
\end{equation}
Here, $\rho$ is the energy density of the fluid which is present 
everywhere. However, the pressure it exerts on the two subspaces
are different in general.

Conservation of energy-momentum
$(T^{\mu}_{\nu;\mu}=0)$ implies 
\begin{equation}
 \frac{d}{dt}(\rho a^3 b^D) \, + \, P_a b^D \frac{da^3}{dt} \, + \, P_b a^3 \frac{db^D}{dt} \, = \, 0 \label{cons}
\end{equation}
For the sake of simplicity, the equations of state are assumed
to yield $P_a = w_a \rho$ and $P_b = w_b \rho$, where $w_a$ and $w_b$ 
are constants.
On solving equation (\ref{cons}), we, then, have
\begin{equation}
 \rho \, = \, \rho_0 \, \frac{a_0^{3(1+w_a)} \, b_0^{D(1+w_b)}}{a^{3(1+w_a)} \, b^{D(1+w_b)}} \label{rhoeq}
\end{equation}
Here, $\rho_0$ is the energy density corresponding to the epoch when
$a = a_0$ and $b=b_0$.  While we could have generalized equation
(\ref{Tmunu}) to include multiple fluids, this would have added little
to our analysis. Hence, we desist from such complication. Indeed, as
we shall see later, the matter content has relatively little importance.

\section{Cosmological Solutions}
\subsection{General features}
The evolution of the Universe is governed by the Einstein's equations 
(with the conservation equations being built into it.) 
As can be seen, we have five variables and six equations, namely the 
three Einstein equations (\ref{Eeq00},\ref{Eeqii},\ref{EeqII}), the 
conservation equation (\ref{cons})  and the 
two equations of states. 
However, once we Eq. \ref{cons} is considered,
only one out of equations (\ref{Eeqii},\ref{EeqII}) is independent. 

We start the evolution at $t = M^{-1}$. While the classical equations
are involved for $t\lesssim M^{-1}$, the particular choice is just one
of the convenient choices and does not have any influence on the
dynamics as a result of time translation invariance.  It will be
convenient to express all the quantities in terms of dimensionless
units, namely
\begin{equation}
\begin{array}{rcl c rcl cc}
\tau &\equiv& \displaystyle t \, M
& \qquad \quad & 
H_a &\equiv& \displaystyle 
    \frac{a'}{a}=\frac{\dot{a}}{a \, M}
\quad \quad \quad H_b &\equiv& \displaystyle 
    \frac{b'}{b}=\frac{\dot{b}}{b \, M}
 \\[2ex]
\bar{\rho} &\equiv& \displaystyle \frac{\rho}{M^{4+D}} 
& & 
\bar{P_a} &\equiv& \displaystyle \frac{P_a}{M^{4+D}} 
\quad \quad \quad \quad \quad \bar{P_b} &\equiv& \displaystyle \frac{P_b}{M^{4+D}}\,
\end{array}
\end{equation}
Here, prime represents derivative with repect to $\tau$.
The initial value of $\tau$ is denoted by $\tau_0$.
Note that for $k_{1,2}=0$, the equations are independent of 
the absolute values of $a(t)$ and $b(t)$ but depend only on the 
values relative to those at present time. This is reflected by the fact 
that $a(t)$ and $b(t)$ enter into the equations(for $k_{1,2}=0$) only 
as logarithmic derivatives.

We can re-express the Equations 
(\ref{Eeq00},\ref{Eeqii},\ref{EeqII}) in terms of dimensionless quantities as 
\begin{eqnarray}
-\bar{\rho} & =& -\lambda \, \left[3 \, H_a^2 
                                  +3 \,  D \, H_a \, H_b 
                                  + \com{2} \, H_b^2 \right]
                 + 12 \, \gamma \, \left[ D \, H_a^3 \, H_b 
                                  + 3 \, \com{2} \, H_a^2 \, H_b^2 
                                  + 3 \, \com{3} \, H_a \, H_b^3 
          + \com{4} \, H_b^4 \right] \label{H0}\\
\bar{P_a} & = &  H_a' \, f_1(H_a, H_b)
                            + H_b' \,  f_2(H_a, H_b)
                + f_3(H_a, H_b)  \label{H1}  \\
\bar{P_b} &=& H_b' \, f_4(H_a, H_b) 
           + H_a' \, f_5(H_a, H_b)  + f_6(H_a, H_b)
\label{H2}
\end{eqnarray}
where
\begin{equation}
\begin{array}{lcl}
f_1(H_a, H_b) & \equiv & \displaystyle
2 \, \left(  
- \lambda + \gamma \left[4 \, D \, H_a \, H_b 
                            + 4 \, \com{2} \, H_b^2 \right]\right) 
\\[1.5ex]
f_2(H_a, H_b) & \equiv & \displaystyle
 -D \,\lambda 
                            + 4 \, \gamma \, \left[ D \, H_a^2 
                            + 3 \, \com{3} H_b^2 
                            + 4 \, \com{2} \, H_a \, H_b \right] 
\\[1.5ex]
f_3(H_a, H_b) & \equiv & \displaystyle
- \lambda \, \left[3 \,H_a^2 
              + \comq{2} H_b^2 
              + 2 \, D \, H_a \, H_b \right]
\\[1ex]
   & + & \displaystyle        \gamma \left[ 8 \, D \, H_a^3 \, H_b 
              - 2 \, D \, (3-5D) \, H_a^2 \, H_b^2 
              + 3 \,(D-1) \, \com{3}  \, H_b^4  
            + 8 \, D \, \com{2} \, H_a \, H_b^3 \right]
\\[1.5ex]
f_4(H_a, H_b) & \equiv & \displaystyle
   -(D-1) \, \lambda 
                             + 12 \, \gamma \left[  \comp{3} \, H_b^2
         +2 \, \comp{2} \, H_a \, H_b 
                             + (D-1) \, H_a^2 \right] 
\\[1.5ex]
f_5(H_a, H_b) & \equiv & \displaystyle
- 3 \,\lambda 
                           + 12 \, \gamma \, \left[ H_a^2 
                           + 2 \, (D-1) \, H_a \, H_b 
                           + \com{2} \, H_b^2 \right] 
\\[1.5ex]
f_6(H_a, H_b) & \equiv & \displaystyle
           - \lambda \, \left[6 \, H_a^2+\com{2} \, H_b^2 
           + 3 \, (D-1) \, H_a \, H_b \right] 
\\[1ex]
 & + & \displaystyle
            12 \, \gamma \, 
    \left[H_a^4             + \com{4} \, H_b^4 
      + (D- 1) \, \left\{3 \,  H_a^3 \, H_b 
           +(2D-3) \, H_a^2 \, H_b^2
           + \comp{2} \, H_a \, H_b^3 \right\}
\right]
\end{array}
   \label{the_f's}
\end{equation}
Note that eqn.(\ref{H0}) is invariant under the joint transformation
$(H_a, H_b) \to (-H_a, - H_b)$. Similarly, 
$f_i(H_a, H_b) = f_i(-H_a, -H_b)$. This is
reminiscent of the analogous property 
of the $(1+3)$--dimensional case. Thus, for an empty universe, 
both expanding ($H_a > 0$) 
and contracting ($H_a < 0$) solutions for the scale factor of the visible 
sector are equally possible and the growth (or otherwise) of the 
universe is determined by the initial conditions. 
We will return to this point later.

\begin{figure}[!htbp]
\centering
% \subfigure[]
{
\includegraphics[width=2.5in,height=2.8in,angle=270]{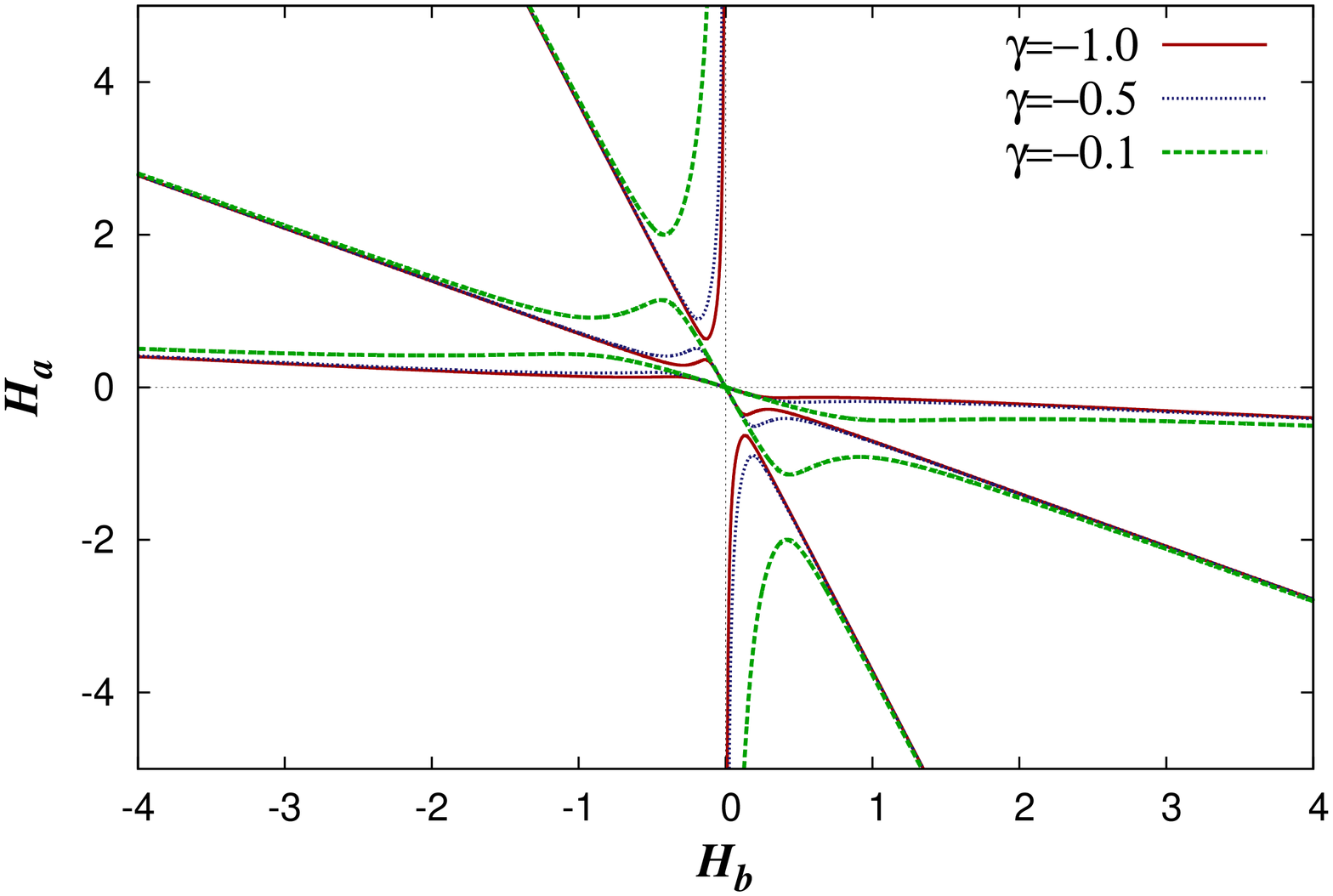}
}
{
\includegraphics[width=2.5in,height=2.8in,angle=270]{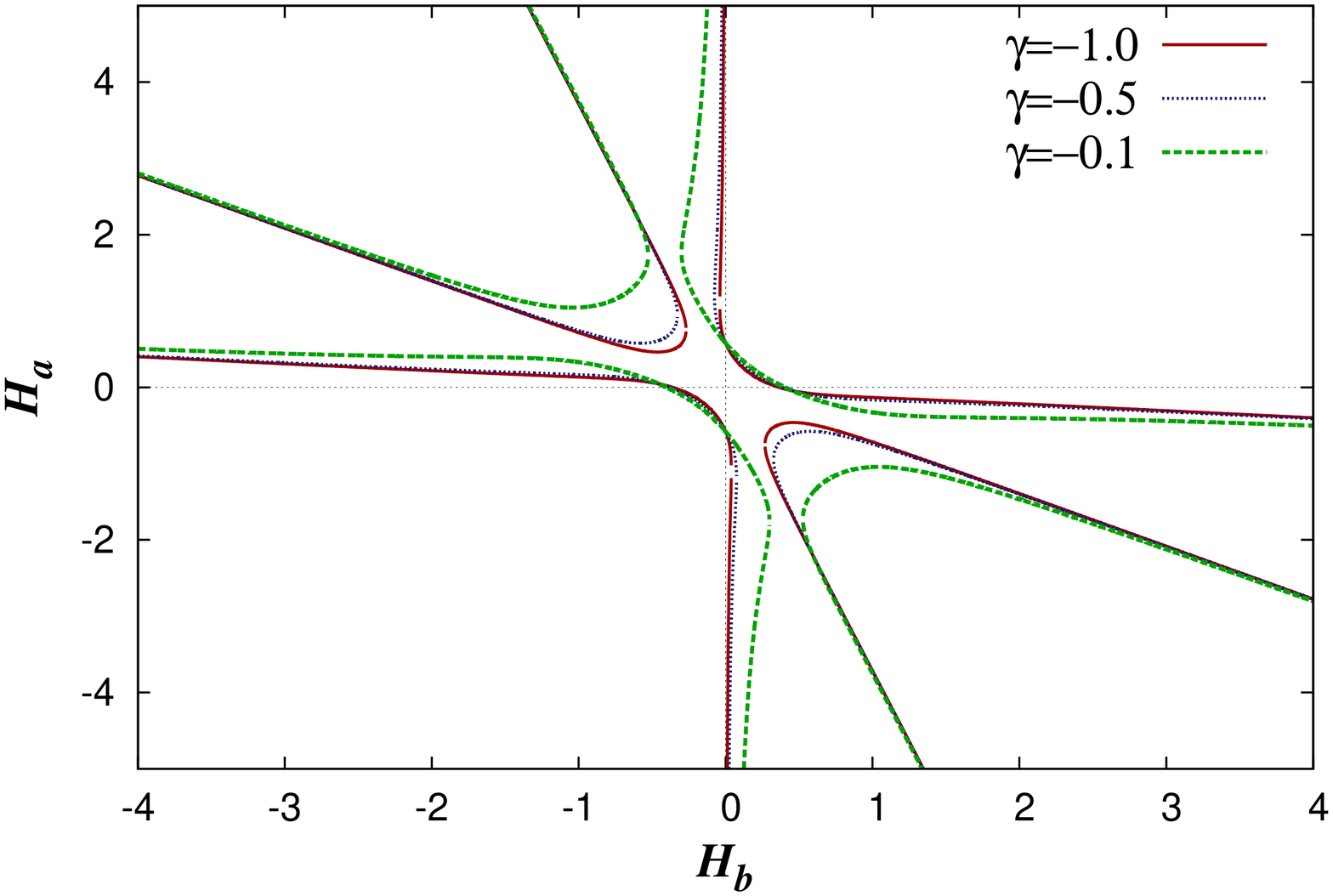}
}
\caption{Curves correspond 
to different $\gamma$ values for $\bar{\rho}=0.0$[Left Panel] and $\bar{\rho}=1.0$[Right Panel].}
\label{fig:hahb_curves}
\end{figure}

It is instructive to examine the structure of eqn.(\ref{H0}) alone, keeping
dynamics aside for the moment.  Being cubic in $H_a$, this leads to
three solutions for a given $(\gamma, \bar \rho, H_b)$
combinations. In other words, there are three branches for the
solution\footnote{While it may seem that with eqn.(\ref{H0}) being
  quartic in $H_b$, there should four branches instead, clearly such
  an argument is faulty. Note that, for $D=2$, there are only two branches}. 
  In Fig.\ref{fig:hahb_curves}, we
depict these branches as plots of $H_a(H_b)$ for specific values
of $\gamma$ and $\bar\rho$. While $\bar\rho = 0$ corresponds to an
empty universe, $\bar\rho = 1$ would indicate a cosmological constant
comparable to the scale of quantum gravity, or in other words, the
limit of applicability of the theory. Thus, during its evolution, the
energy density of the universe must lie somewhere in between these
limits.  Depending on the values of $(\gamma, \bar \rho, H_b)$, the
solutions for $H_a$ could either be all real, or one real and a
complex-conjugate pair.  Complex solutions are, of course, not
admissible on physical grounds, and this is indicated in
Fig.\ref{fig:hahb_curves} by the fact of there being only one solution
for certain values of $H_b$.

Irrespective of the nature of the dynamics, it is obvious 
that, for a given $\bar\rho(t)$, the pair $(H_b, H_a)$ must 
lie on a curve that is bounded by the curves for $\bar\rho=0$ and $\bar\rho=1$. The extent to which they can 
move away from a specific curve is governed by the 
evolution of $\bar\rho(t)$. As Fig.\ref{fig:hahb_curves} shows,
this dependence on $\bar\rho$ can be significant, especially 
for small $(H_a, H_b)$. However, if for some reasons, such a configuration is 
reached only at late times, this dependence would turn out to be 
irrelevant. Even a large primordial $\bar\rho$ would be seen to be 
diluted to insignificant amounts. 

Before we solve for the dynamics, it is useful to dwell upon a
few more issues:
\begin{itemize}
\item Note that we can never have
  both $H_{a,b} > 0$ for $\bar{\rho}=0$. This is a
  consequence of the signs of $\gamma$ and $\lambda$, 
  dicated as they are by other physical considerations. 
  This, though is not true for $\bar{\rho}>0$. However, since
  $\bar{\rho}$  quickly decreases in magnitude to almost a 
  vanishing value, if the visible universe  
  is expanding at a given epoch, the hidden dimensions are contracting
  and vice versa. 

\item The above result has a striking consequence. 
  Unless the universe started 
with a configuration opposite to what we have today---namely a large 
$b(\tau = 1)$ 
and a small $a(\tau = 1)$---we must have had a relatively small 
$|H_b(\tau)|$ during the entire history of the universe. In particular, 
the inflationary epoch should have had $H_a \gg |H_b|$.
This immediately tells us which of the branches in Fig.\ref{fig:hahb_curves}
should the initial reside on.

\item It is easy to appreciate that a large value for $|H_b|$ in the
  present epoch would be inadmissible as this would lead to the
  effective cosmological parameters changing rapidly. Moreover, since
  the theory is applicable only for $b(\tau) > 1$, the contraction of
  the hidden dimensions must have essentially stopped at some distant
  past\footnote{This conclusion can be evaded only if the universe had
    started with an immensely large $b(\tau)$.}. Since $H_b \approx 0$
  implies that $ H_a^2 = \bar\rho / 3\lambda$ (a result that one would
  have obtained in standard $(1+3)$--dimensions as well), we must have
  $\lambda > 0$ as we have already discussed.

\end{itemize}

Since it is possible to choose initial conditions such that
$|H_b(\tau = 1)| \ll H_a(\tau = 1)$, it is now clear (courtesy
Fig.\ref{fig:hahb_curves}) that this hierarchy would be maintained.
This has the immediate consequence that $\bar\rho(\tau)$ would be a
fast decreasing function of $\tau$ (see eqn.\ref{rhoeq}) as long as
$w_a > -1$ and $w_b$ is not very large\footnote{In other words, the 
contraction in $b(\tau)$ should not offset the expansion in $a(\tau)$ 
in the evolution of $\bar\rho(\tau)$.}. Since this 
inequality is satisfied by all ordinary matter,
it is quite apparent that the universe quickly settles down to a phase
where the matter density (or the pressure thereof) has little bearing on the evolution of scale factors.
This result\footnote{Note that we have not assumed this simplification
  in our actual calculations.}  helps us understand better the assymptotic nature of the solutions to the
dynamical equations that we now turn to, in particular, the lack of
sensitivity to the matter content.

\begin{figure}[!htbp]
\centering
% \subfigure[]
{
\includegraphics[width=2.5in,height=2.8in,angle=270]{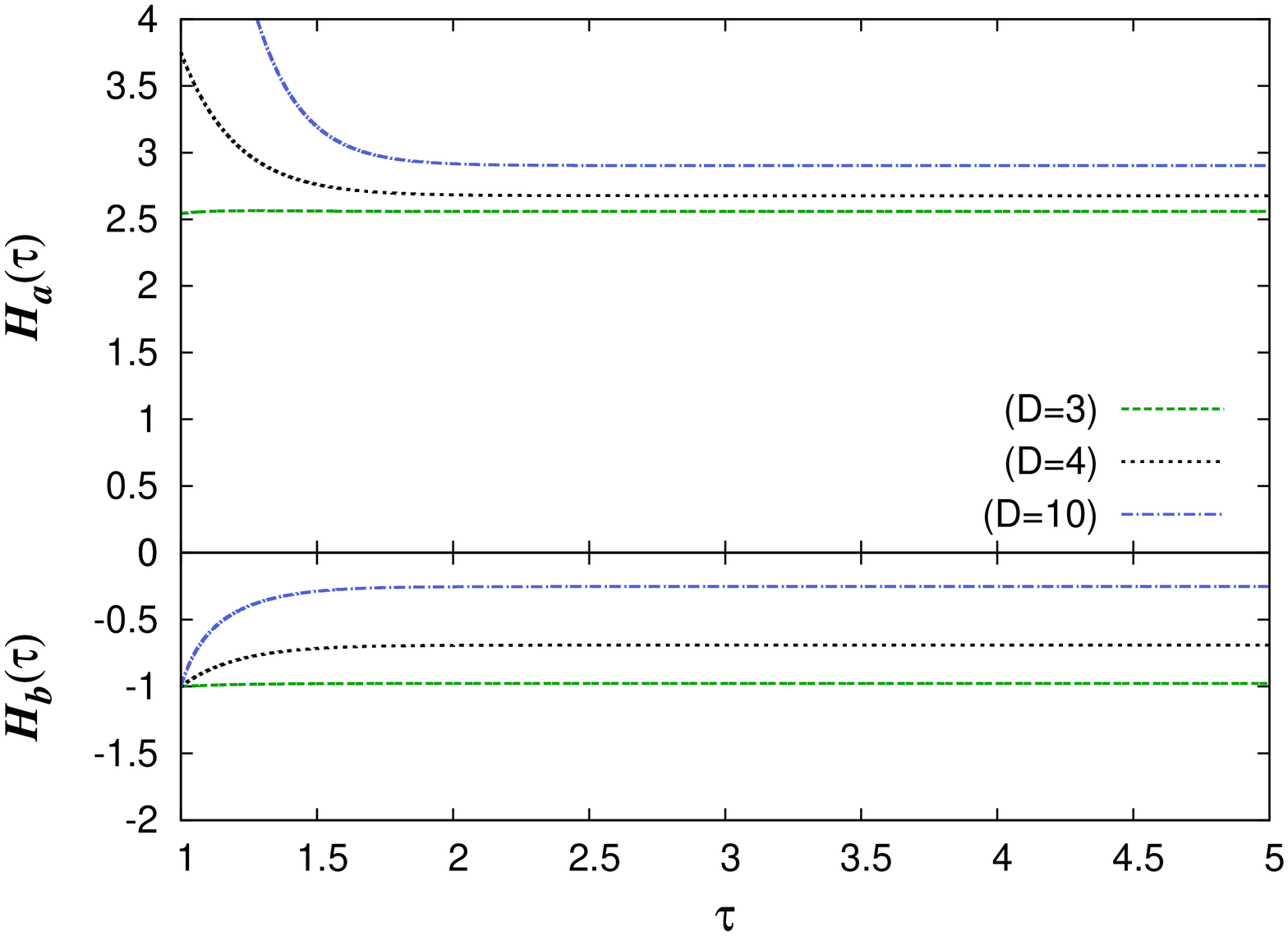}
}
% \subfigure[]
{
\includegraphics[width=2.5in,height=2.8in,angle=270]{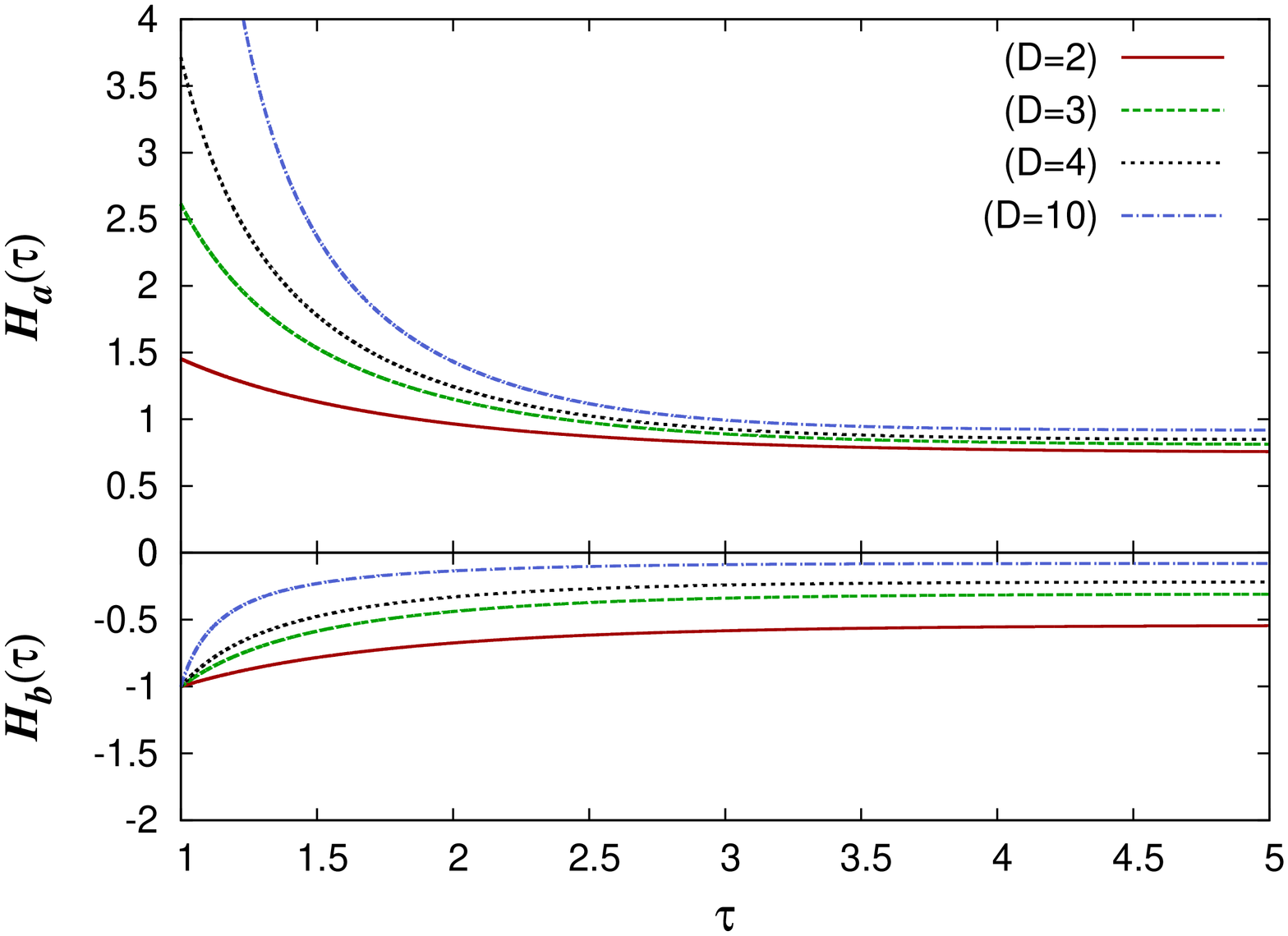}
}
\caption{Behaviour of the $H_a$ and $H_b$ with the rescaled time $\tau$
for $\gamma = -0.1$ [left panel] and $\gamma = -1.0$ [right panel]. 
In either case, we have chosen the parameters $w_a=1.0/3.0$, $w_b=0.0$, 
i.e. a pure radiation confined to $(1 + 3)$-dimensions, 
and the initial  conditions $\bar\rho(\tau = 1) = 1$, $H_b(\tau = 1)= -1.0$.}
\label{fig:hahb}
\end{figure}

\begin{figure}[!htbp]
\centering
% \subfigure[]
{
\includegraphics[width=2.5in,height=2.8in,angle=270]{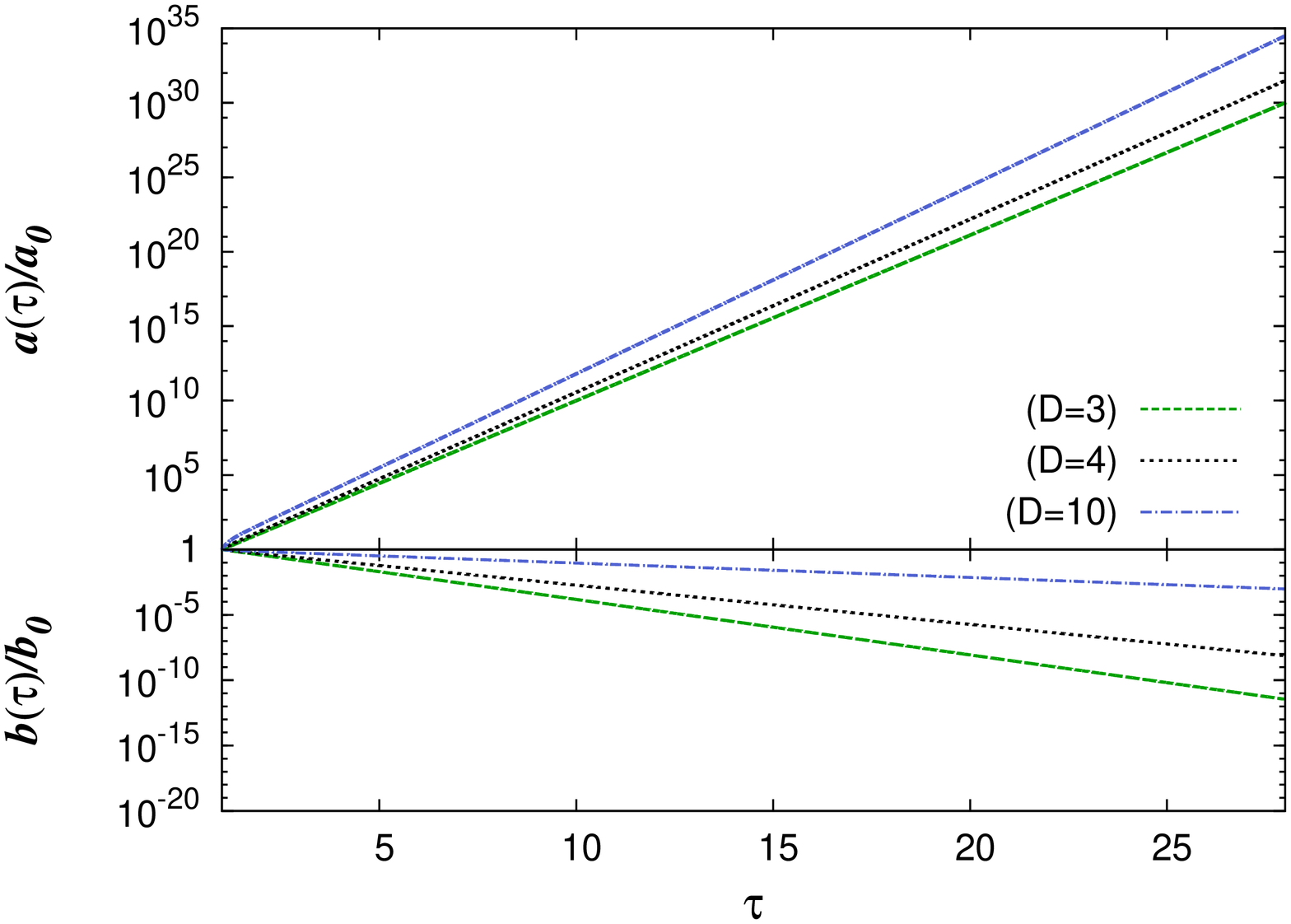}
}
% \subfigure[]
{
\includegraphics[width=2.5in,height=2.8in,angle=270]{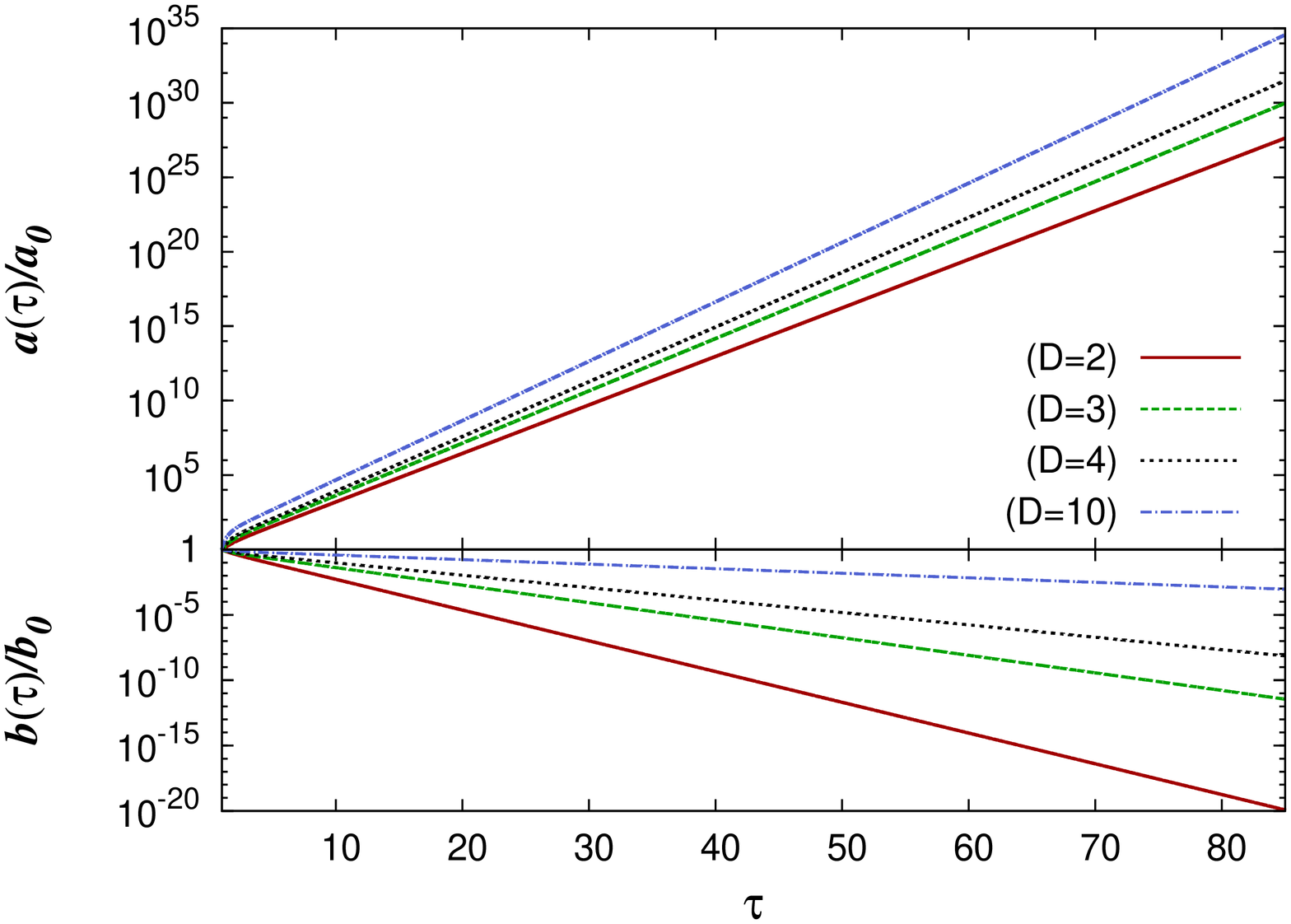}
}
\caption{Behaviour of the normalized scale factors $a(\tau) $ and $
  b(\tau)$ with the rescaled time $\tau$ for $H_b(1.0)=-1.0,
  w_a=1.0/3.0$ and $w_b=0.0$.  The left and right panels correspond to
  $\gamma = -0.1$ and $\gamma = -1.0$ respectively.}
\label{fig:ab}
\end{figure}

For a linear equation of state (as we have assumed), we can 
easily eliminate $\bar \rho$ and $\bar P_a$ by combining 
eqs.(\ref{H0}\&\ref{H1}). With $H_a = H_a(H_b)$, what
remains to be done is to solve a single first-order non-linear
differential equation\footnote{One could also have solved
  eqns.(\ref{H1} \& \ref{H2}) for $H'_{a,b}$ and, thereby obtained not
  only $d H_a / d H_b$, but also $H_a(t)$ and $H_b(t)$. The two
  procedures are equivalent.}. 
In Fig.\ref{fig:ab}, we illustrate the
Hubble parameters  for certain specific choices of the constants 
$\gamma, D$ and the initial condition $H_b(\tau_0) = -1$. In each 
case the particular branch of the solutions $H_a(H_b)$ has been 
chosen so as to lead to acceptable amount of inflation. Note that, 
in each case, $H_b$ initially increases from $H_b(\tau_0) = -1$, only 
to reach a fixed point. Courtesy eqn.(\ref{H0}), a similar fate befalls 
$H_a$ too, with it {\em decreasing} to a {\em positive} fixed point value. 
Increasing $D$ has the twin effect of increasing  
$|H_a^{\rm asymp}|$ (decreasing $|H_b^{\rm asymp}|$), 
while delaying the epoch of settling at the fixed point. 
The dependence on 
$|\gamma|$ is analogous to that on $D$.
This can be appreciated by realizing that the effect of the Gauss-Bonnet
term becomes more pronounced with increase in $D$ and decrease in $|\gamma|$.

It is useful to consider, here, the senstivity of the
evolution to the parameters $w_{a,b}$ as well as the initial condition
$\bar\rho(\tau_0)$.  Let us consider the latter point first. As we have
argued above, a comparison of the two panels of
Fig.\ref{fig:hahb_curves} demonstrates that there is little difference
between $\bar\rho = 1$ and $\bar\rho = 0$. On the other hand,
naturalness demands that $\bar\rho \lapp 1$. Thus, even though
changing $\bar\rho(\tau_0)$ does have a bearing on the evolution, the
effect is discernible only for a small time scale (the transient
region) and quickly becomes insignificant. For precisely the same
reason, the effect of $w_{a,b}$ is marginal too, as long as that the
energy density does not remain unchanged from a significant starting
value, (i.e., unless $w_{a,b}$ conspire so that $\bar\rho(\tau) \sim
{\cal O}(1)$ over a significant range $\tau \gg \tau_0$).

\vskip 10pt

A constant positive $H_a$ (negative $H_b$) implies exponentially inflating 
visible universe (deflating hidden dimensions). 
In Fig.\ref{fig:ab}, we illustrate the
resultant scale factors. As mentioned earlier,
all evolution is independent of the initial value of the scale factors. 
Understandably, at late times $a(\tau)$ continues to increase 
with the time $\tau$, while $b(\tau)$ continues to decrease\footnote{We 
postpone the discussion of monotonicity at {\em all times} $\tau$ until 
later.}. Note, though, that $a(\tau)$ cannot be inflating forever. 
Nor can $b(\tau)$ decrease continually, for that would either imply 
a current value of $b(\tau)$ to be smaller than the scale of quantum gravity 
(a regime beyond the validity of our treatment) or a very large 
initial value  $b(\tau_0)$, which, though allowed in principle, is 
aesthetically unpleasing. Thus the goal should be a large value 
of the ratio $R_{\rm end} \equiv a(\tau_{\rm end}) / b(\tau_{\rm end})$, where 
$\tau_{\rm end}$ denotes the epoch of the end of inflation. Without going 
into the mechanism of ending inflation, it is clear from  Fig.\ref{fig:ab}
that $R_{\rm end}$ increases monotonically\footnote{This could have been gleaned 
from Fig.\ref{fig:hahb} as well.} with $D$, and, thus, a larger 
$D$ proffers, in some sense, a better solution.

\subsection{Asymptotic dynamics}

The numerical solutions obtained in the preceding section suggest
that, after a transient behaviour, the evolution, at late times
asymptotes to exponential growth (decay) for the visible (hidden)
sector.  We examine, now, whether this feature is a generic one or is
it specific to a class of initial conditions and/or parameters.

\begin{figure}[!htbp]
\centering
% \subfigure[]
{
\includegraphics[width=3.2in,height=4.0in,angle=270]{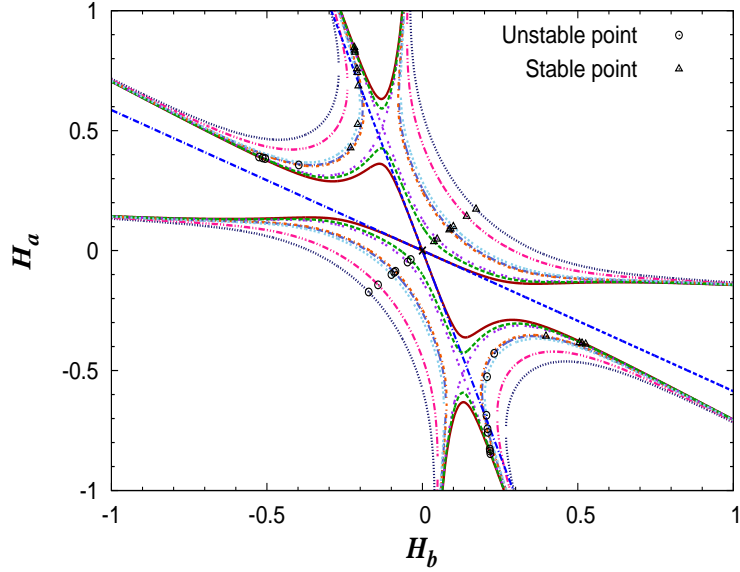}
}
\caption{Phase space plot of $H_a$ and $H_b$ for $D=4$. 
Curves in this figure correspond to $\bar{\rho}=0.0$(solid line)
and $\bar{\rho}=1.0$(dotted line) for $\gamma=-1.0$.}
\label{fig:hahb_D4}
\end{figure}

The existence of exactly exponential solutions for both $a(\tau)$ and
$b(\tau)$ would, of course, be contingent on having $H_a' = H_b' = 0$.
Now, $H'_{a,b}$ can be algebraically solved for (in terms of
$H_{a,b}$) using eqns.(\ref{H1}\&\ref{H2}).
Vanishing derivatives of hubble parameters, 
thus, translate to curves in the $H_a$--$H_b$ plane (each
corresponding to specific values of $\bar P_a$ and $\bar P_b$).  The
overlap of these curves with those corresponding to the constraint
equation (\ref{H0})---see, for example,
Fig.\ref{fig:hahb_curves}---albeit for an allowed combination $\left(
\bar\rho(\tau), \bar P_a(\tau), \bar P_b(\tau) \right)$ would, then,
indicate the fixed points. While this entails solving the full set of
the dynamical equations, it is instructive to consider a special
choice, namely a constant $\bar \rho(\tau)$. For the linear equation
of state that we have assumed, this essentially implies a cosmological
constant\footnote{It is interesting to note the possibility of having
a solution with $w_{a,b} \neq 1$, but with the Hubble parameters
satisfying $a^{3 \, (1 + w_a)} \, b^{D \, (1 + w_b)} \approx {\rm
constant}$, at least asymptotically.}. In Fig.\ref{fig:hahb_D4}, we
illustrate these fixed points for a particular configuration, namely
$(D= 4, \ \gamma = -1.0)$.

Not all the fixed points would be stable. Stability, in this context,
implies that time evolution from the neighbourhood of this point should bring $H_{a,b}$ {\em to} this point
rather than take it away from it. In other words, for a small
displacement from the fixed point, the derivatives $H'_{a, b}$ should
point towards the point. In Fig.\ref{fig:hahb_D4}, small triangles
(circles) denote the stable (unstable) fixed points. Note that each
stable point $(\bar H_a, \bar H_b)$ is paired with an unstable one at
$(-\bar H_a, -\bar H_b)$. This pairing can be understood by recalling
that $f_i(H_a, H_b) = f_i(-H_a, -H_b)$ (see eqn.\ref{the_f's}), and,
hence the derivatives must satisfy $H'_{a,b}(\{H_{a,b}\}) = -
H'_{a,b}(\{-H_{a,b}\})$.  Once a stable fixed point $(H_a = \alpha,
H_b = \beta)$ has been reached, the system would continue in that
state, leading to exponential growth (decay) thereon.

At this stage, it is imperative to distinguish the case of the
cosmological constant from the more general scenario of an evolving
$\bar \rho(\tau)$. Assuming a non-trivial fixed point $(\alpha, \beta)
\neq (0,0)$ exists, the former must satisfy eqn.(\ref{H0}). On the
other hand, the matter energy density (other than that for a
cosmological constant) changes with time. This, immediately,
demonstrates that no such fixed point can exist in the generic case.
Nonetheless, as we argued earlier, for $w_{a} > -1$, the matter
density quickly falls to insignificant levels (as long as we do not
have $w_b \gg -1$), and thus, for such matter content, the asymptotic
behaviour would be almost indistinguishable from that due to an empty
universe.

%%%%%%%%%%%%%%%%%%%%%%%%%%%%%%%%%%%%%%%%%%%
\begin{center}
\begin{table}[!htpb]
\begin{tabular}{|c|c||c|c|c||c|c|c|}
\hline
\multicolumn{2}{|c|}{} & \multicolumn{3}{|c|}{Stable Solutions}&\multicolumn{3}{|c|}{Unstable Solutions}\\
\hline
$D$ & $\gamma$ & $\alpha$ & $\beta$ & $\delta = \beta/\alpha$ & $\alpha$ & $\beta$ & $\delta = \beta/\alpha$\\
\hline \hline
$2$ & $-1.0$ &$0.75$ & $-0.54$ & $-0.72$ & $-0.75$ & $0.54$ & $-0.72$\\
\hline
 &$-0.5$&$1.06$&$-0.77$&$-0.72$&$-1.06$&$0.77$&$-0.72$\\
 \hline \hline
$3$&$-1.0$&$0.81$&$-0.31$&$-0.38$&$-0.81$&$0.31$&$-0.38$\\
   &      &$-0.31$&$0.81$&$-2.61$&$0.31$&$-0.81$&$-2.61$\\ 
   \hline
   &$-0.5$&$1.14$&$-0.44$&$-0.38$&$-1.14$&$0.44$&$-0.38$\\
   &      &$-0.44$&$1.14$&$-2.59$&$0.44$&$-1.14$&$-2.59$\\
   \hline
   &$-0.1$&$2.56$&$-0.98$&$-0.38$&$-2.56$&$0.98$&$-0.38$\\
   &      &$-0.98$&$2.56$&$-2.61$&$0.98$&$-2.56$&$-2.61$\\
\hline \hline
$4$ & $-1.0$ & $0.85$ & $-0.22$ &$-0.26$& $-0.85$ & $0.22$ &$-0.26$\\
    &        & $-0.39$ & $0.52$ & $-1.33$& $0.39$ & $-0.52$ & $-1.33$\\
\hline
 & $-0.5$ & $1.19$ & $-0.31$ &$-0.26$& $-1.19$ & $0.31$ &$-0.26$\\
 &        & $-0.55$& $0.74$ & $-1.34$ & $0.55$& $-0.74$ & $-1.34$\\
\hline
 \hline
 &$-0.1$&$2.67$&$-0.69$&$-0.26$&$-2.67$&$0.69$&$-0.26$\\
 &      &$-1.23$&$1.66$&$-1.35$&$1.23$&$-1.66$&$-1.35$\\
 \hline
 \hline
$10$ &$-1.0$&$0.92$&$-0.08$&$-0.09$&$-0.92$&$0.08$&$-0.09$\\
     &      &$-0.59$&$0.22$&$-0.37$&$0.59$&$-0.22$&$-0.37$\\  
 \hline 
  &$-0.5$&$1.298$&$-0.11$&$-0.09$&$-1.298$&$0.11$&$-0.09$\\
  &      &$-0.83$&$0.31$&$-0.37$&$0.83$&$-0.31$&$-0.37$\\
 \hline 
  &$-0.1$&$2.9$&$-0.25$&$-0.09$&$-2.9$&$0.25$&$-0.09$\\
  &      &$-1.85$&$0.69$&$-0.37$&$1.85$&$-0.69$&$-0.37$\\
 \hline \hline
\end{tabular}
\caption{Different analytic solutions for $\rho=0$}
\label{table1}
\end{table}
\end{center}

%%%%%%%%%%%%%%%%%%%%%%%%%%%%%%%%%%%%%%%%%%%

Reverting to the cosmological constant, it is clear that, for $D=3$,
the symmetry between the visible and the hidden sectors guarantees
that every (un)stable fixed point $(\alpha, \beta)$ would be
accompanied by another of the same ilk, namely $(\beta, \alpha)$. For
$D > 4$, on the other hand, no such symmetry is present, and we would,
in general have two rather different stable points (each with its
unstable mirror image). For $D = 2$, the aforementioned three solution
branches collapse to two, and the two stable fixed points merge. These
features have been exhibited in Table~\ref{table1}. 

For a given $D$, if $\gamma$ is changed, the asymptotic values
$(\alpha, \beta)$ do change, but the ratio $\delta \equiv \beta /
\gamma$ remains unchanged (see Table~\ref{table1}). This can be
understood by noting that, for an empty universe,
eqns.(\ref{H0}--\ref{H2}), can be trivially recast in terms of the
scaled variables $\sqrt{-\gamma} \, H_a$ and $\sqrt{-\gamma} \,
H_b$. The very same property dictates that the values $\alpha$ and
$\beta$ are proportional to $1 / \sqrt{|\gamma|}$. If we denote the 
epoch of reaching the fixed point by $\tau_f$, then, at a later time
$\tau > \tau_f$, we have
\begin{equation}
\frac{a(\tau)}{b(\tau)} = \frac{a(\tau_f)}{b(\tau_f)} \;
                     \exp\left[(\alpha - \beta) \, (\tau - \tau_f) \right] \label{relative_expansion}
\end{equation}
Thus, to maximize the growth of the ratio
of the scale factors, we need to maximize the
difference $(\alpha - \beta)$. Clearly, this is aided by decreasing
$|\gamma|$. While this may seem paradoxical at first, this is not so
given the fact that the limit $\gamma \to 0$ is not
straightforward and has to be taken with care\footnote{Note
that $\gamma$ (or, equivalently $\epsilon$) essentially introduces a
new scale in the theory. For example, the aforementioned scaling of
$H_{a,b}$ could, instead, be understood in terms of rescaling time by
the same factor of $\sqrt{-\gamma}$.}. This is demonstrated by 
Table~\ref{table1}. 

The dependence of $(\alpha, \beta)$ on the number of hidden dimensions
$D$ is more subtle. Analytically, it is best understood in the limit
of very large $D$. Approximating, say, $\com{n} \approx D^n / n!$ {\em
etc.}, the pure-gravity version of eqns.(\ref{H0}--\ref{H2}) can now
be recast in terms of $(H_a, \widetilde H_b \equiv D \, H_b)$ with all
dependence on $D$ being absorbed in $\widetilde H_b$. This immediately
implies that $\beta$ should scale as $D^{-1}$ while $\alpha$ should be
free of any $D$-dependence. Of course, such a scaling is exact only in the
infinite-$D$ limit, and corrections to this behaviour should be apparent
for moderate $D$ values. A perusal of Table~\ref{table1} show, nonetheless,
that $\beta \propto D^{-1}$ is a fairly good approximation, while the
relative variation in $\alpha$ is much smaller. Indeed, even these residual
dependences can be understood analytically if a $1/D$ expansion is performed.
For our purposes, though, such an exercise is of limited use and we shall
desist from it at the current juncture. Once again, considerations
analogous to those of eqn.(\ref{relative_expansion}) tells us that
a larger $D$ value is more suitable for effecting large inflation with
a moderate contraction of the hidden dimensions. Once again, 
Table~\ref{table1} clearly demonstrates this. 

\subsection{End of Inflation}

One issue that we have not addressed so far
is the mechanism to end inflation. In different models
of extra-dimensions motivated inflationary
scenarios\cite{Brown,panda,david,guido,john}, the issue of the end of
inflation and reheating is addressed in different ways. As the scale
factor of the extra-dimension decreases, it is expected that a stage is reached when
quantum gravity effects become important.  Our analysis, however, is
of classical nature, and, hence, is not 
equipped to address this issue directly. Several speculations can be 
made though. For one, it is conceivable that quantum gravity itself would 
generate a pressure that would arrest such an interminable contraction, 
thereby forcing $a(t)$ to come out of the inflationary phase to a FRW-like one.

While this may seem to be a dissatisfactory solution, 
let us consider if such a situation can be reached dynamically 
with some (relatively) well-understood physics input. If $b(t)$ is to reach a constant value, 
we must impose $H_b=0$ and $H_b'=0$ in eqns.(\ref{H0},\ref{H1},\ref{H2}).
A consistent solution needs 
\begin{equation}
 \bar{\rho} \, = \, -\frac{3}{2} 
\, \bar{P_a}
+\frac{\sqrt{  9 -96 \gamma \bar{P_b}+72 \gamma \bar{P_a}  ( 2 \bar{P_a} \gamma  + 3  )  }}{8 \gamma} \ .
\end{equation}
Of course, such an equation of
state seems extremely unnatural and contrived. Moreover, assuming this
to be an identity would not admit the dynamics that we have seen so
far, and, thus, would negate the very reason for our analysis. Rather,
what we need is a equation of state that, at some suitably late epoch,
smoothly flows to the above or at least to a close approximation
thereof\footnote{The approximation has to be good enough so that 
$b(\tau)$ does not change appreciably.}. Even this flow seems a bit contrived 
if one considers only a single matter component. However, on the inclusion 
of different fields, with differing kinetic terms and potentials, such an
eventuality in the evolution of the total stress-energy tensor 
can be managed with relative ease.

Yet another possibility arises if we treat the coefficient of the 
Gauss-Bonnet term ($\epsilon$) not as a constant but a dynamical quantity. 
This is not an unreasonable assumption for the Gauss-Bonnet term and is supposed 
to have arisen as the result of quantum corrections. Once $\epsilon$ is assumed 
to be a field, it will, naturally, evolve with time. If the potential for $\epsilon$ 
be such that $\epsilon = 0$ is a (local) minimum, the evolution equations
would naturally force the visible universe to come out of the inflating 
phase to a  decelerating era.
This issue and other 
associated questions such as reheating and the generation of 
density perturbations will be addressed in a subsequent work.

% At this point, we would like to point out that we do not have a neat way 
% to end inflation. We assume that the inflation ends when the
% size of $b$ becomes comparable to $1+3$ 
% dimensional planck scale. After that, solutions enters into the 
% usual higher  dimensional FRW case which gives standard power-law solutions.
% The analysis of the mechanism for the graceful exit will be addressed 
% in a future paper.

\section{Conclusions and Discussions}
In the context of a higher dimensional cosmology, we have, in this 
paper, given a mechanism to produce inflation using the Gauss-Bonnet 
term which provides a natural extension of the Einstein-Hilbert action. 
Unlike several common models of inflation where a field (called the inflaton field) 
is required for an inflationary phase \cite{robert,robert1,shinji,sriramkumar} in the 
Early Universe, in our case, we have eliminated the necessity of such a field. 
The universe goes into an inflationary phase 
just from the interplay of the extra dimensions and the Gauss-Bonnet 
term in the action.
 
The parameter space characterized by the the expansion rate in the
normal and extra dimensions, $H_a$ and $H_b$ , exhibits a rich
structure with a variety of possible solutions depending on the
initial conditions.  There are both stable and unstable fixed points.  
If, for simplicity, linear equations of state are
assumed, then the matter density dies down fast with time, as long as
$w_a > -1$ and $w_b \leq 1$. Thus, asymptotically, the system reaches
the limit of vacuum solution, which in our case turns out to be
exponential behaviour of scale factors with time.  More importantly,
if one of the scale factors increase with time, the other
decreases. This is a very desirable situation as we would like the
normal dimensions to expand while the extra dimension should become
smaller. A large expansion is achievable for a significantly
wide range of the dimensionless Gauss-Bonnet parameter
$\gamma$. Indeed, a $\gamma$ small enough to be consistent with a
quantum-mechanical origin is admissible on this count. And while the
asymptotic values of $H_a$ and $H_b$ do depend on the Gauss-Bonnet
parameter, their ratio is essentially insensitive to it. For a given
value of $\gamma$, the expansion rate of the visible dimensions
($H_a$) depends only weakly on the number of extra dimensions, $D$.
On the other hand, $H_b$ has a stronger dependence on $D$. In fact,
for large $D$, $H_b$ is roughly inversely proportional to $D^{-1}$.

The end of inflation, the generation of density perturbations and
reheating are some of the open questions in this context. While we
have proferred some possible solutions to the first of these, a
detailed mechanism fulfilling all constraints needs to be worked out.

\acknowledgments{IP
  acknowledges the CSIR, India for assistance under grant
  09/045(0908)/2009-EMR-I. DC thanks the Department of Science and Technology,
  India for assistance under the project DST-SR/S2/HEP-043/2009 and
  acknowledges partial support from the European Union FP7 ITN
  INVISIBLES (Marie Curie Actions, PITN-GA-2011-289442).   TRS thanks CSIR, India for assistance under the project
  Ref O3(1187)/11/EMR-II.
    Authors acknowledge the facilities provided
  by the Inter University Center For Astronomy and Astrophysics, Pune,
  India through the IUCAA Resource Center(IRC), University of Delhi, New Delhi, India.}


\newpage
\begin{thebibliography}{99}
\bibitem{KK1} J. M. Overduin and P. S. Wesson, \, {\it Kaluza-Klein gravity}, \,
{\it Phy. Reports} \, {\bf 283} \, (1997) \, 303
\bibitem{KK2} E. Witten, \, {\it Search for a realistic Kaluza-Klein Theory}, \, {\it Nucl. Phy. B} \,
{\bf 186} \, (1981) \, 412
\bibitem{RS1} L. Randall and R. Sundrum, \, {\it Large Mass Hierarchy from a Small Extra Dimension}, \,
{Phys. Rev. Lett.} \, {\bf 83} \, (1999) \, 3370
\bibitem{RS2} L. Randall and R. Sundrum, \, {\it An Alternative to Compactification}, \,
{Phys. Rev. Lett.} \, {\bf 83} \, (1999) \, 23
\bibitem{ADD} N. Arkani-Hamed, S. Dimopoulos and G. Dvali, \, {\it The hierarchy problem and new dimensions at
a millimeter}, \, {\it Phys. Lett. B} \, {\bf 429}  \, (1998) \, 263
\bibitem{maeda} K. Maeda and T. Torii, \, {\it Covariant gravitational equations on a brane world
with a Gauss-Bonnet term}, \, {\it Phys. Rev. D} \, {\bf 69} \, (2004) \, 024002
\bibitem{Lovelock} D. Lovelock, \, {\it The Einstein Tensor and Its Generalizations}, \,
{\it J. Math. Phys.} \, {\bf 12} \, (1971) \, 498
\bibitem{naresh} N. Dadhich, \, {\it On the Gauss-Bonnet Gravity}, \, arXiv:hep-th/0509126v3
\bibitem{paul} B. C. Paul and S. Mukherjee, \, {\it Higher-dimensional
cosmology with Gauss-Bonnet terms and the cosmological-constant problem}, \,
{\it Phy. Rev. D} \, {\bf 42} \, (1990) \, 2595
\bibitem{sami} R. Chingangbam et al, \, {\it A note on the viability of Gauss-Bonnet Cosmology},
\, {\it Phys. Lett. B}, \, {\bf 661} \, (2008) \, 162
\bibitem{madore} J. Madore, \, {\it Cosmological applications of the Lanczos Lagrangian}, \,
{\it Class. and Quan. Grav.} \, {\bf 3} \, (1986) \, 361
\bibitem{boulware} D. G. Boulware and S. Deser, \, {\it String-Generated Gravity Models}, \, 
{Phys. Rev. Lett.} \, {\bf 55} \, (1985) \, 2656 
\bibitem{andrew} K. Andrew, B. Bolen and C. A. Middleton, \, 
{\it Solutions of higher dimensional Gauss-Bonnet FRW cosmology}, \,
{\it General Relativity and Gravitation} \, {\bf 39} \, (2007) \ 2061
\bibitem{toporensky} D. M. Chirkov and A. V. Toporensky, \, {\it On stability of power-law
solution in multidimensional Gauss-Bonnet cosmology}, \, arXiv:1212.0484
\bibitem{Brown} A. R. Brown, \, {\it Boom and Bust Inflation: a Graceful Exit via
Compact Extra Dimensions}, \, {\it Phys. Rev. Lett.}, \, {\bf 101} \,
(2008) \, 221302
\bibitem{panda} S. Panda, M. Sami and I. Thongkool, \, {\it Reheating the D-brane 
universe via instant preheating}, \, {\it Phys. Rev. D}, \, {\bf 81} \, (2010) \,103506
\bibitem{david} D. Gherson, \, {\it Constraints on the size of the extra dimension from
Kaluza-Klein gravitino decay}, \, {\it Phys. Rev. D}, \, {\bf 76} \, (2007) \, 043507
\bibitem{guido} G. D'Amico et al, {\it Inflation from Flux Cascades}, \, 
arXiv: 1211:3416
\bibitem{john} J. H. Brodie and D. A. Easson, {\it Brane inflation and reheating}, \,
{\it JCAP}, \, {\bf 12} \, (2003) \, 004
\bibitem{robert} R. H. Brandenberger, \, {\it Cosmology of the Early Universe}, \, 
arXiv: 1003.1745
\bibitem{robert1} R. H. Brandenberger, \, {\it A Status Review of Inflationary Cosmology}, \,
arXiv: hep-ph/0101119
\bibitem{shinji} S. Tsujikawa, \, {\it Introductory review of cosmic inflation}, \,
arXiv: hep-ph/0304257
\bibitem{sriramkumar} L. Sriramkumar, \, {\it An introduction to inflation and 
cosmological perturbation theory}, \, arXiv:0904.4584
\end{thebibliography}
\end{document}